# A SLIDING MODE LATERAL VELOCITY OBSERVER

H. E. Tseng
Ford Research Laboratory

Scientific Research Lab. 2101 Village Road, Dearborn, MI 48124 USA
Phone : +(313) 845-5667
Fax : +(313) 248-5167
E-mail : htseng@ford.com

Abstract: A lateral velocity estimation scheme whose stability can be analytically derived (rather than empirically demonstrated through cut-and-try) is attempted. The designed adaptive sliding mode observer shows robust performance under a wide variety of maneuvers/ environments, including the more challenging slow J-turn on low mu.

## 1. INTRODUCTION

Lateral velocity of a vehicle, and the corresponding vehicle body side slip angle and tire side slip angles, are important indicators of the vehicle handling operating region and tire-road conditions, especially during limit handling. Many recent active electronic chassis control systems use side slip angle as a control reference [1,2,3,4]. The lateral velocity of a production vehicle, however, is not directly measured due to practical issues such as cost, accuracy, and reliability in current sensor technologies. Therefore, the estimation of vehicle lateral velocity is a vital issue and has been widely discussed in the literature.

Senger and Kortum in 1989 developed a vehicle handling model and tire model to estimate the lateral speed [5]. Their approach assumed that the tires were operating in the linear region with known cornering stiffness. That assumption, of course, does not hold in real applications.

Cao in 1994 used a combined parameter and state estimation approach to accommodate unknown and varying cornering stiffnesses [6]. However, the effectiveness of this method without yaw rate measurement is doubtful based on our experience and the method is unlikely to overcome road bank disturbances.

Farrelly and Wellstead, in 1996, proposed both a physical modeling approach and a kinematics modeling approach [7]. Their physical modeling based approach is limited for real life application while their kinematics approach has some potential.

Kaminaga and Naito, in the 1998 AVEC conference in Nagoya, Japan, presented a Lyapunov based adaptive observer [8]. They attempt to devise a methodology with 'a high degree of mathematical stability'. The performance shown on paper is impressive but leaves out the discussion of practical issues such as road bank disturbances.

Liu and Peng, in a 1998 ASME Journal article, presented an identification scheme for simultaneous state and parameter estimation to overcome the unknown vehicle parameters [9]. However, it is observed that the methodology requires one to two cycles of vehicle maneuvers before the estimation converges for non-nominal conditions.

Fukada in 1998 AVEC presented a combined observer/direct integration method [2]. The method strived to balance the robustness towards modeling error and the robustness towards signal bias. Much of its logic is empirically derived. This method, later refined by Nishio et al. [10], is understood to be the key logic of latral velocity estimation of Toyota's Electronic Stability Control (ESC) system.

Anthony Zanten discussed their active yaw control system, Bosch VDC, in an 2000 SAE paper [3]. He utilized the relation between (longitudinal) slip stiffness and (lateral) cornering stiffness during large longitudinal slip conditions to obtain a more reliable estimate of the slip angle. However, this methodology has to be combined with an integration method to compensate for less reliable operating conditions such as during free rolling.

Hac and Simpson, in a 2000 SAE paper, presented a combined method similar to Toyota's approach but with a full state observer and more analysis [4]. However, it appears to resort again to empirically derived logic when road bank disturbances are considered.

Overall, it appears that the analytical methodologies in the literature do not provide robustness [11] against practical issues such as road disturbances or the time-varying nature of vehicle cornering stiffnesses while the empirically derived methodologies fall short in stability analysis.

In this paper, a model based adaptive observer utilizing the sliding mode techniques will be discussed, its mathematical stability shown and its robust performance under a wide variety of realistic environments verified.

## 2. VEHICLE MODEL

To develop a model-based observer, a bicycle model with two degrees of freedom is chosen. With $\delta$ denoting the steering wheel angle, $u$ as the longitudinal speed, $v$ as the lateral speed, $r$ as the yaw rate, $m$ as the vehicle mass, $I$ as the yaw moment of inertia, $a$ and $b$ as the distance from vehicle c.g. to the front and rear axles, the state space model can be represented as

$$\begin{bmatrix}\dot{v}\\ \dot{r}\end{bmatrix}=\begin{bmatrix}-\frac{c_f+c_r}{mu} & -u+\frac{-ac_f+bc_r}{mu}\\ \frac{-ac_f+bc_r}{Iu} & \frac{-a^2c_f-b^2c_r}{Iu}\end{bmatrix}\begin{bmatrix}v\\ r\end{bmatrix}+\begin{bmatrix}\frac{c_f}{m}\\ \frac{ac_f}{I}\end{bmatrix}\delta$$

where $c_f$ and $c_r$ are the unknown front and rear tire cornering stiffnesses.

The longitudinal speed and steering wheel angle are measured and the output measurements of the vehicle model include yaw rate and lateral acceleration.

## 3. ADAPTIVE OBSERVER WITH REDUCED STATE/AUXILIARY DERIVATIVE FEEDBACK

### 3.1 Design of an Adaptive Observer

Consider a plant that can be described as a linear model whose output can be specified with a reduced set of the system states, i.e.

$$\dot{x}=Ax+Bu$$
$$y=Cx+Du$$

where $C=[0_{n_o\times(n-r)},C_{n_o\times r}]$, $x\in R^{n\times 1}$, $y\in R^{n_o\times 1}$, and $r<n$.

Assuming ($A$, $C$) is an observable pair, no structural uncertainty and the parameter uncertainty resides only in the model's state dynamics (i.e. matrices $A$ and $B$) but not in the output measurement (i.e. matrices $C$ and $D$).

If the derivatives of the states which the output is statically independent of can be measured and are bounded, then the following observer scheme is suggested.

$$\begin{aligned}\dot{\hat{x}}&=\hat{A}\hat{x}+\hat{B}u+L(y-C\hat{x}-Du)\\&=W(\hat{x},u)\hat{\theta}+L(y-C\hat{x}-Du)\end{aligned}\quad(1)$$

$$\dot{\hat{\theta}}=\Gamma^{-1}W^T(\hat{x},u)P\tilde{x}'\quad(2)$$

where $\Gamma>0, P>0$, $\tilde{x}=x-\hat{x}$ and

$$\tilde{x}'=\begin{bmatrix}\tilde{x}'_{n-r}\\ \tilde{x}_r\end{bmatrix},\qquad \tilde{x}'_{n-r}(0)=0\quad(3)$$

$$\dot{\tilde{x}}'_{n-r}=-\Lambda\tilde{x}'_{n-r}+(dx_{n-r}/dt)-(d\hat{x}_{n-r}/dt)\quad(4)$$

$$\Lambda=\begin{bmatrix}\lambda_1 & 0 & 0\\ 0 & \ddots & 0\\ 0 & 0 & \lambda_{n-r}\end{bmatrix}\qquad \lambda_i>0,\forall i\quad(5)$$

where $x_{n-r}\in R^{(n-r)\times 1}$ is the collection of the states that the output is statically independent of.

Conceptually, the adaptive law (Eq. 2) feeds back a 'virtual measurement error' where the 'virtual measurement' of the un-measurable state is denoted as $x'$ if we define $x'=\tilde{x}'+\hat{x}$.

Claim:

If *(A-LC)* is stable for all possible $A$, $\tilde{x}(0)=0$, and $\hat{A},\hat{B}$ are bounded, then there exist a finite time $t$ where the suggested adaptive observer can ensure the bound of $\tilde{x}(t)$ as a function of $\lambda$ which can be reduced with the reduction of $\lambda$ if $\hat{x}(\tau)$ is bounded $\forall\tau\in[0,t]$.

Proof:

With Eq. (1), the error dynamics can be described as

$$\begin{aligned}\dot{\tilde{x}}&=(A-LC)\tilde{x}+\tilde{A}\hat{x}+\tilde{B}u\\&=(A-LC)\tilde{x}+W(\hat{x},u)\tilde{\theta}\end{aligned}\quad(6)$$

Define $V=\tilde{x}^TP\tilde{x}+\tilde{\theta}^T\Gamma\tilde{\theta}$, where $P$ and $\Gamma$ are positive definite, we have

$$\begin{aligned}\frac{dV}{dt}&=\tilde{x}^T[(\hat{A}-LC)^TP+P(\hat{A}-LC)]\tilde{x}\\&\quad+2\tilde{\theta}^T[W^T(x,u)P\tilde{x}+\Gamma\dot{\tilde{\theta}}]\end{aligned}\quad(7)$$

Using the adaptation law suggested by Eq. (2), and define $f$ as the virtual measurement error, i.e., $f=x'-x=\tilde{x}'-\tilde{x}$, we have

$$\begin{aligned}\frac{dV}{dt}&=-\tilde{x}^T((\hat{A}-LC)^TP+P^T(\hat{A}-LC))\tilde{x}\\&\quad+2\tilde{\theta}^T[\tilde{W}^TP\tilde{x}-W^T(\hat{x},u)Pf]\\&=\tilde{x}^T((A-LC)^TP+P^T(A-LC))\tilde{x}-2\tilde{\theta}^TW^T(\hat{x},u)Pf\\&=-\tilde{x}^TQ\tilde{x}-2\tilde{\theta}^TW^T(\hat{x},u)Pf\end{aligned}$$

where $(A-LC)^TP+P(A-LC)=-Q$ and $Q>0$.

For a finite time $t$, we have $f(t)\to 0$ if $\lambda_i\to 0,\forall i$, or $0<|f(t)|<\varepsilon(\lambda)$. Since the first term, $-\tilde{x}^TQ\tilde{x}$, does not depend on $\lambda$, $\tilde{x}(t)$ is bounded and its bound can be reduced with the reduction of $\lambda$ since $\tilde{\theta}(\tau)$ and $\hat{x}(\tau)$ are bounded for $\forall\tau\in[0,t]$.

### 3.2 Discussions of the Designed Observer

#### 3.2.1 Practical Implementation

While the above claim seems very limited with its assumptions in the boundedness of $\tilde{\theta}$ and $\hat{x}$ and the initial condition of $\tilde{x}(0)=0$, these assumptions may not be an issue for practical problems such as the lateral velocity estimation problem at hand. The lack of asymptotic convergence at infinite horizon is not a practical issue since our actual problem deals with a time-varying plant and some extra mechanisms with fast convergence properties (within finite time) are to be expected.

#### 3.2.2 Stability Analysis

Using the Lyapunov function, Kaminaga and Naito have derived an adaptive observer that guarantees the convergence of lateral velocity estimate [8]. Unfortunately, the derived adaptive law is not practical since it requires full state information. In their actual implementation, they first reduced the feedback requirement from estimation error to the sign of estimation error with the sliding mode technique. They then circumvented the difficulty by introducing 'a substitute for beta (i.e. side slip angle)' which is the time derivative of beta. Although they argued that "the difference between the estimated value of beta and its real value due to changes in cornering power would be approximately the same", their Lyapunov analysis becomes irrelevant to their implemented methodology.

In our adaptive observer design discussed above, a 'virtual measurement' is introduced and the stability of the algorithm to be implemented is illustrated.

3.2.3 Virtual Measurement

In our observer design, a 'virtual measurement' is constructed when the state is not measurable but its derivative is. This segment discussed the insight regarding its construction. For example,

$$\dot{\tilde{x}}' = -\lambda\tilde{x}' + \left(\dot{x}\,|_m - \dot{\hat{x}}\right)$$
$$= -\lambda\tilde{x}' + (\dot{\tilde{x}} + d)$$

where $x'$ is the virtual measurement, $\tilde{x}' = x' - \hat{x}$ is the virtual estimation error and $d$ is the measurement noise. We have

$$\tilde{x}' = \frac{s}{(s+\lambda)}\tilde{x} + \frac{1}{(s+\lambda)}d \qquad (8)$$

and

$$f = x' - x = \tilde{x}' - \tilde{x} = \frac{-\lambda}{(s+\lambda)}\tilde{x} + \frac{1}{(s+\lambda)}d$$

We see that the virtual measurement error at time *t, f(t),* can be made arbitrarily small as $\lambda \to 0$ if $\tilde{x}(0) = 0$, $\tilde{x}$ does not have any low frequency content and *d=0.*

However, any uncompensated low frequency noise in *d* gets amplified as $\lambda$ decreases (see Eq. 8). Therefore, a trade-off decision has to be made for the design parameter $\lambda$ between the attenuation of measurement noises and the integrity keeping of $\tilde{x}$.

Note also the virtual measurement can be expressed as a combination of the high frequency component of true signal and the low frequency component of estimated value (as well as some attenuated disturbances.)

$$x' = \frac{s}{(s+\lambda)}x + \frac{\lambda}{(s+\lambda)}\hat{x} + \frac{1}{(s+\lambda)}d$$

Therefore, the virtual measurement, *x'*, can be made close to the true signal, *x*, as long as the estimate, $\hat{x}$, captures the low frequency content while keeping the measurement noises at a minimum.

For our specific problem and bicycle model, the state $x=[v, r]^T$ is composed of lateral velocity *v* and yaw rate *r*. Yaw rate can be directly measured from a gyro rate sensor while there is no economical sensor available for lateral velocity. The time derivative of lateral velocity, however, can be measured from lateral accelerometer, $a_y$, yaw rate, *r*, and vehicle speed, *u*, as

$$\dot{v} = a_y - r \cdot u + g \sin\varphi .$$

We see that the lateral velocity derivative measured contains road disturbances in addition to electrical noises. Since these noises are inevitable and sometimes substantial, it is impractical to integrate the lateral velocity derivative for a lateral velocity.

While the complete removal of *d* is very difficult, if not impossible, the removal of the more disturbing low frequency content in *d* can be performed by independently estimating road disturbances as shown by Tseng [12]. The virtual measurement suggested further attenuate sensor noises to avoid high frequency errors or drifting issues without unnecessarily washing out the estimated value to zero (which a leaky integrator scheme would do).

3.3 Sliding Mode Application

The above analysis leads to the following design suggestions.

1. To find an *L* such that *A-LC* is stable for all possible *A*
2. To project $\hat{A}$ and $\hat{B}$ such that $\tilde{A}, \tilde{B} \in L_\infty$
3. Make a trade-off decision for the value of $\lambda_i$.
4. Apply sliding mode control methodology to facilitate fast convergence to the sliding surface, $\tilde{x}' = 0$. That is,

$$\dot{\hat{\theta}} = \Gamma^{-1} W^T(\hat{x}, u) P sign(\tilde{x}')$$

and the stability analysis can be adapted by defining a Lyapunov function as $V = p_1|\tilde{x}_1| + p_2|\tilde{x}_2| + \tilde{\theta}^T\Gamma\tilde{\theta}$.

## 4. SIMULATION RESULTS

The suggested sliding mode adaptive observer was implemented against a linear time-invariant bicycle model to study its nominal performance. In this setup, there is no structural uncertainty and the sensor measurements are ideal. The observer robustness against vehicle non-linearities, time-varying parameters, sensor noises, and disturbances will be studied in the section of experimental verifications.

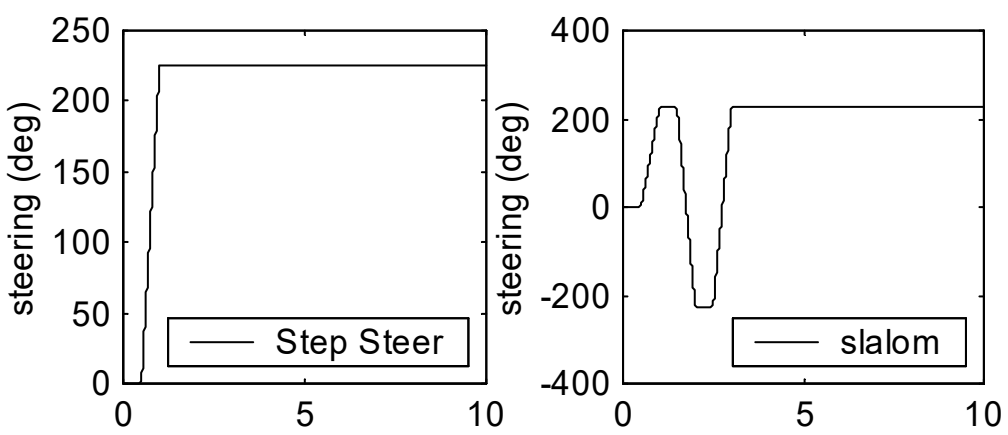

Fig. 4.1 Steering Input of Two Simulated Events

A step steer and a slalom and turn event were studied (Fig. 4.1). We see that with λ being small, the estimation tracked the true value quite well for both events although there is an undesirable high frequency chattering noise due to the sliding mode design. With the sliding mode design, the estimation was quickly pulled to the sliding surface, $\tilde{x}' = 0$. And with λ being small, the virtual measurement is close to the true value ($x' \approx x$) since $\tilde{x}' \approx \tilde{x}$.

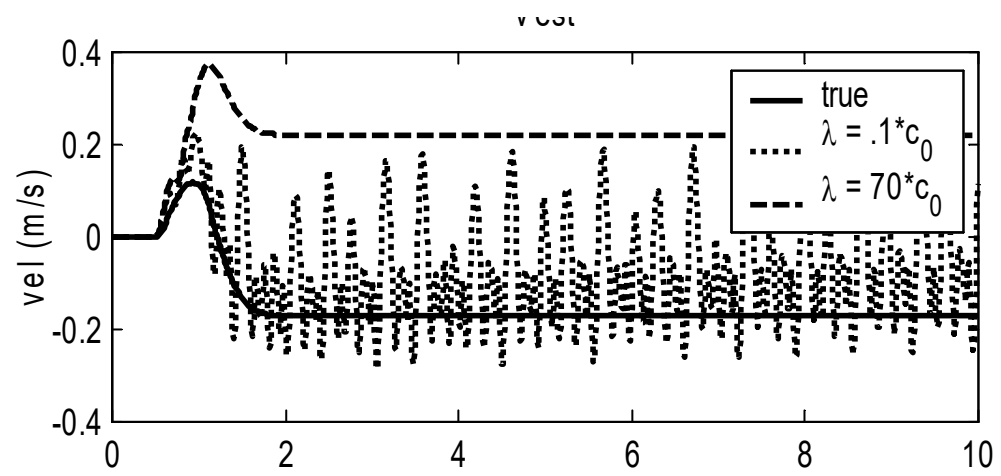

Fig. 4.2 Lateral Velocity Estimate During Step Steer

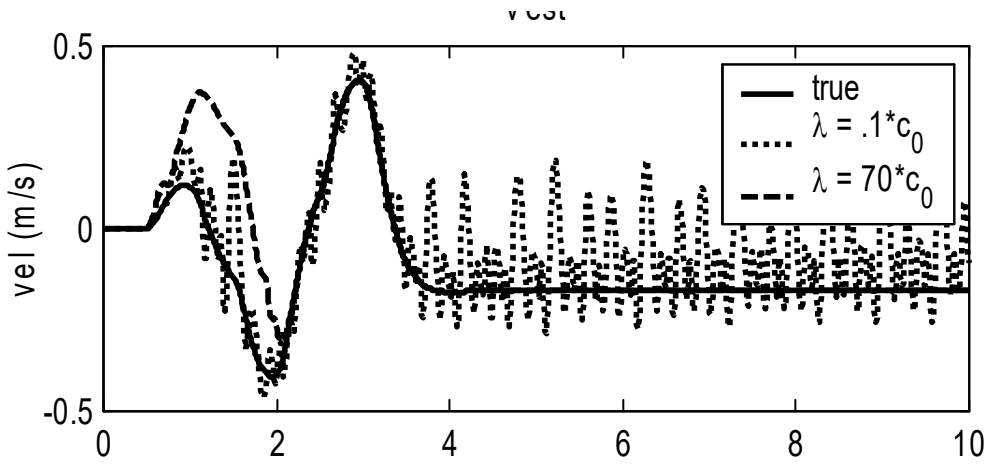

Fig. 4.3 Lateral Velocity Estimate During Slalom

The estimation results using a large $\lambda$ was also studied. We see the estimation error and the parameters may converge after persistent excitation while the performance was poor before parameter convergence (as shown by the large steady state error of the step steer event in Fig. 4.2 and the significant transient error during the slalom event in Fig. 4.3). With $\lambda$ being large, the observer essentially ignores the auxiliary state derivative information and performs similar to an output (or reduced state) feedback adaptive observer at best. An example of an output feedback adaptive observer was presented by Liu and Peng [9] and was found to have some practical issues for direct vehicle application due to its slow convergence rate [11].

Fig. 4.4 shows the convergence of parameter estimates during the slalom event.

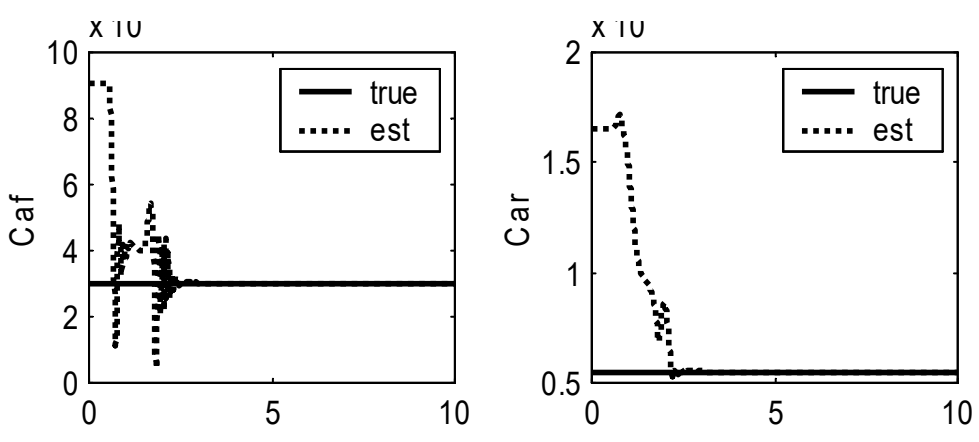

Fig. 4.4 Parameter Convergence During Slalom

## 5. EXPERIMENTAL VERIFICATIONS

The performance of proposed adaptive sliding mode observer under realistic environments and maneuvers is investigated using post-process simulation with vehicle test data. The test maneuvers include lane change, slalom, and J-turn, and the road surfaces tested include asphalt, packed snow, and ice. The sensor signals collected and fed into the sliding mode observer are typically available on an Electronic Stability Control (ESC) equipped vehicle. They are steering wheel angle, yaw rate, lateral acceleration, and longitudinal vehicle speed. The measured and estimated rear axle side slip angles are compared since the rear tire side slip angle is a more transparent indicator of vehicle limit handling than lateral velocity itself or body side slip angle.

The performance is further compared with another well-developed estimation methodology, a kinematics approach developed by Farrelly and Wellstead [7]. This methodology is chosen due to its robustness with regards to model uncertainties and will be referred to as the F&W method [11]. Considering the sliding mode observer is a model-based observer, the kinematics approach is expected to be a competitive contender, if not the limit-performance setter, in cases of relatively road (bank) disturbance free environment (such as the low mu maneuvers/surfaces tested.) It should be mentioned that the Farrelly and Wellstead's kinematics approach has been adapted by integrating their methodology with a physical model based observer to avoid un-observability and the drifting issues during near-zero yaw rate conditions. The F&W observer gain has been tuned to achieve a balance between its sensitivity with respect to noises in longitudinal acceleration measurement and that with respect to noises in lateral acceleration measurement. It was further enhanced by removing possible lateral acceleration bias contained in the sensor measurement with a previously-solved, road-tested bias estimation methodology developed by Tseng [12]. To compare the two methodologies under the same sensor set typically available on an ESC equipped vehicle, the longitudinal acceleration input required by F&W method is emulated with longitudinal velocity derivative. This eliminates the noise of road pitch bias while introducing other sensor noises. Details of the adaptation of F&W method can be found in the paper by Ungoren, Peng, and Tseng [11].

Table 5.1 List of Maneuvers

| Maneuver | Surfaces | Peak Rear Side Slip Angle |
|---|---|---|
| Slow J-turn | low mu | NA |
| Wide Lane Change | low mu | 10 deg |
| Continuous Lane Change | low mu | 20 deg |
| Slalom on Bank | hi mu | 5 deg |

For the reader's reference, Fig. 5.1, 5.4, 5.7, and 5.10 illustrate each tested maneuver listed in Table 5.1 with its model reference yaw rate, $r_{swa}$, (which is a function of vehicle speed and steering wheel angle), measured yaw rate, $r_{meas}$, lateral acceleration measured, $a_y$, estimated lateral acceleration bias (for compensation purpose), *bias*, vehicle traces (as bird's view), and longitudinal speed.

Fig. 5.1 shows a slow drift maneuver on snow. In this maneuver, the driver first steered the vehicle to about an 18 deg/sec and 0.4 g steady state turn, he further manipulated the vehicle to slowly spin the vehicle despite the cut-back in steering. We see that the vehicle yaw rate and the steering (or its corresponding model yaw rate) starts to deviate at around 6 sec into the maneuver while the lateral acceleration is essentially unchanged. Lateral velocity and side slip angle at the rear axle increases between 6 sec. and 10 sec. Beyond 10 sec, the driver applied the brake to stop the vehicle since stability was already lost.

This maneuver, in which the vehicle drifts side-ways with a slow side slip angle derivative (in particular, of about 2 deg/s between 4 to 6 sec. and about 5 deg/s between 6 to 8 sec. into the maneuver), can be considered as one of the benchmark maneuvers in evaluating the performance of lateral velocity estimations. As described by Fukada in his paper of Toyota's effort in slip angle estimation, "To estimate the slip angle accurately in such a slow (J-turn) motion in the nonlinear region is very difficult … considering the influence of road slant or any other factors" [2]. (Note the J-turn in Fukada's paper has a body side slip angle build-up rate of about 9 deg/sec.) Hac and Simpson also confirmed the difficulty in one of their experimental test data, "This (a slow J-turn on low mu) is a difficult maneuver from the estimation viewpoint, because of extremely slippery surface and low speed" [4].

We see the lateral velocity estimate derived from our adaptive sliding mode observer tracked the optical measurement nicely. While both results may be acceptable, the kinematics approach generated less than desirable magnitude between 6 and 8 sec. At 8 sec. into the maneuver, it underestimated the side slip angle (at the rear axle) by 50% or 7 deg while the actual is about 15 deg .

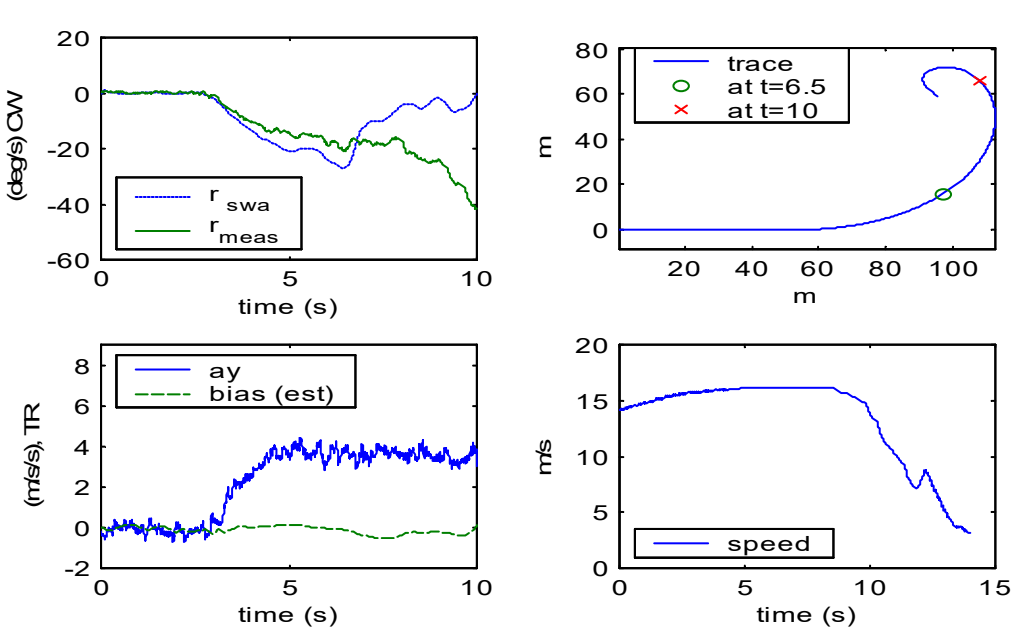

Fig. 5.1 Case 1: Slow J-turn/Side Drift on Snow

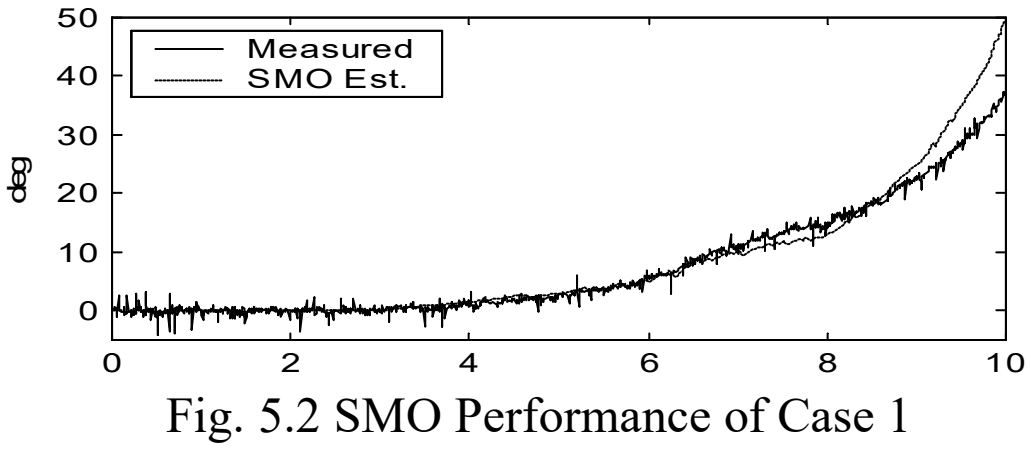

Fig. 5.2 SMO Performance of Case 1

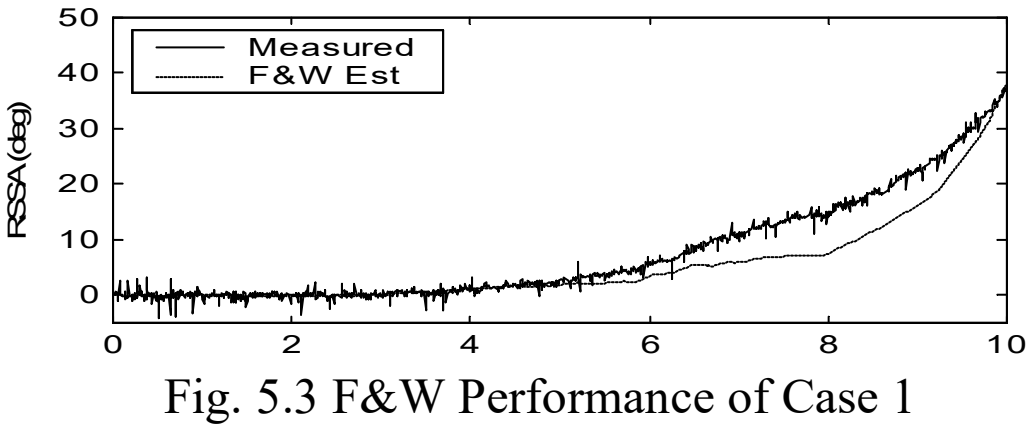

Fig. 5.3 F&W Performance of Case 1

Fig. 5.4 shows a wide lane change maneuver on a low mu surface. The non-linearity can be seen from the deviation between measured yaw rate and modeled yaw rate (calculated from steering wheel angle). The performance of Sliding Mode Observer (SMO) estimate is demonstrated in Fig. 5.5 while the over and under estimate of side slip angle by the F&W method beyond 5 sec. into the maneuver is illustrated in Fig 5.6.

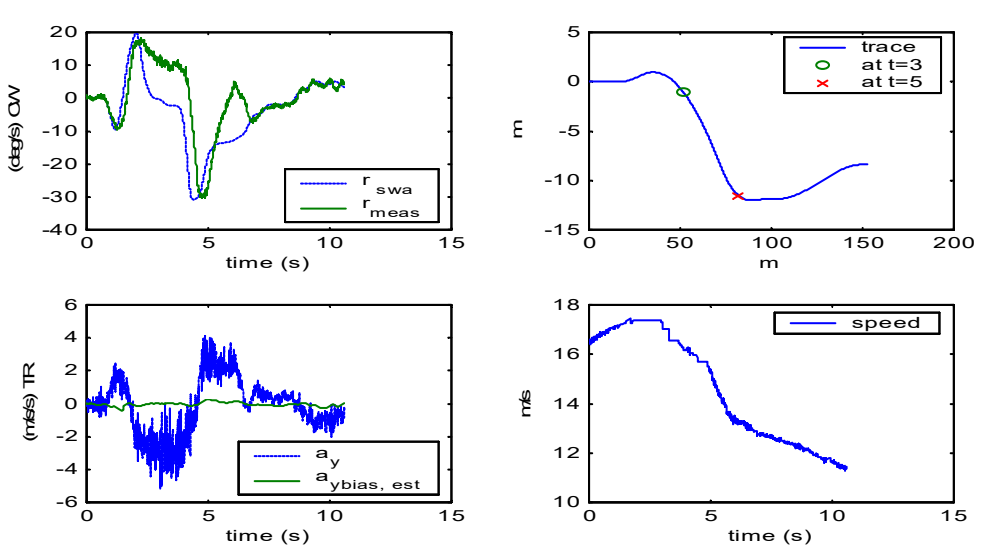

Fig. 5.4 Case 2, Wide Lane Change on low mu

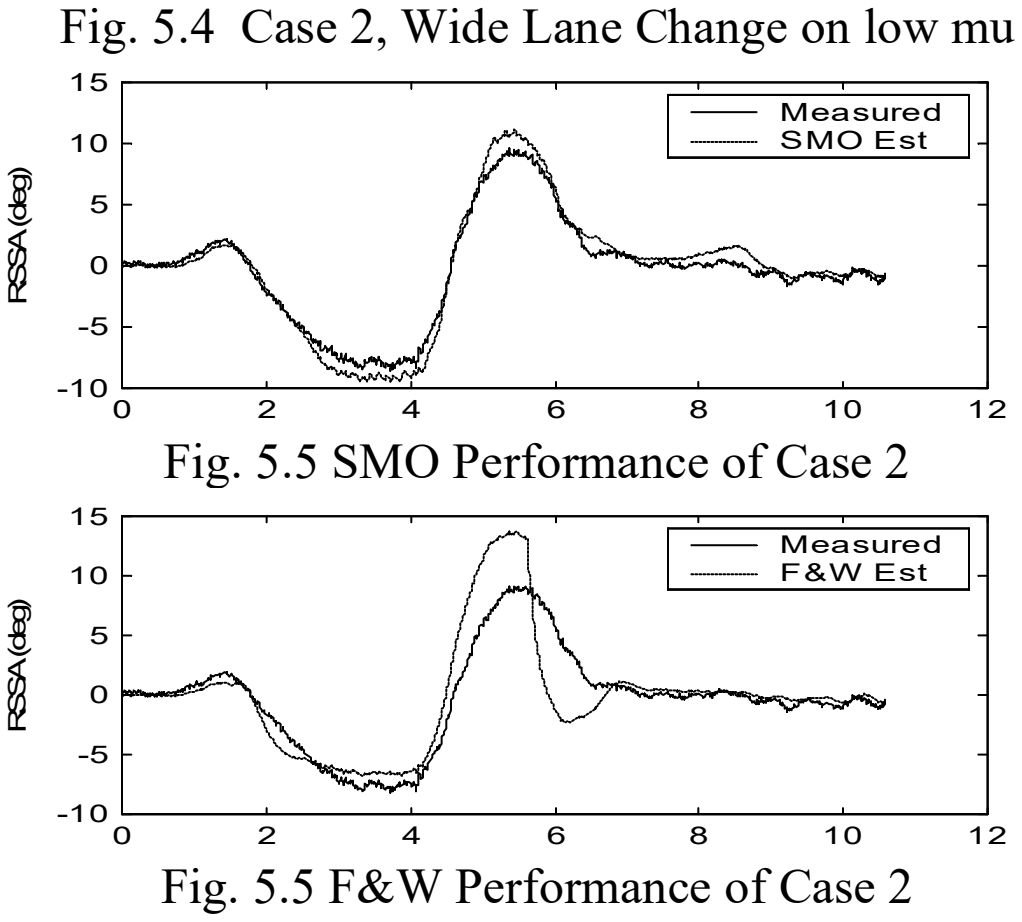

Fig. 5.5 SMO Performance of Case 2

Fig. 5.5 F&W Performance of Case 2

Fig. 5.6 shows a double lane change on packed snow followed by a wide lane change. Fig. 5.7 shows the SMO estimate tracked the measurement well, in both magnitude and phase. Fig. 5.8 revealed a weakness of in the F&W method that results in an undesirable phase shift during a continuous maneuver.

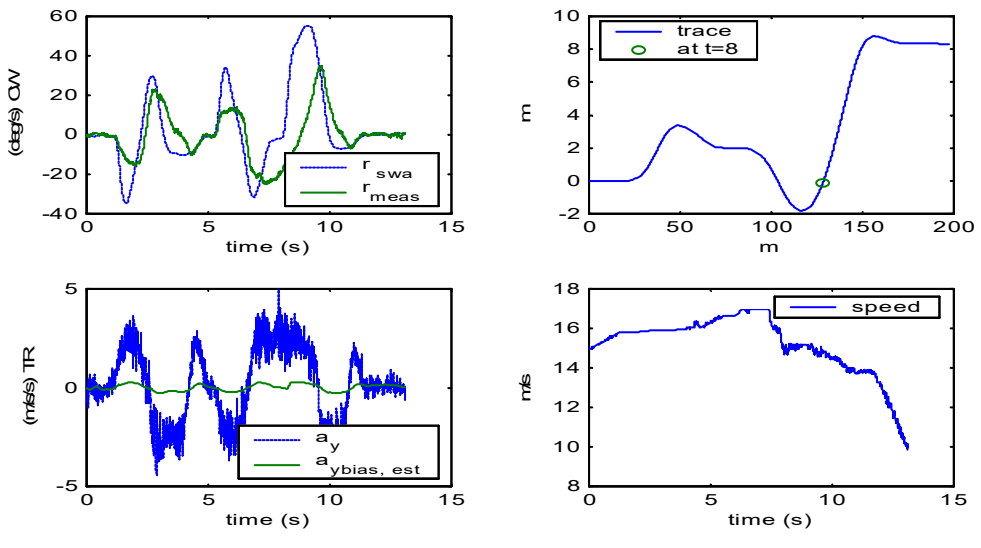

Fig. 5.6 Case 3 Lane Changes on Packed Snow

Measured
SMO Est
RSSA (deg)

Fig. 5.7 SMO Performance of Case 3

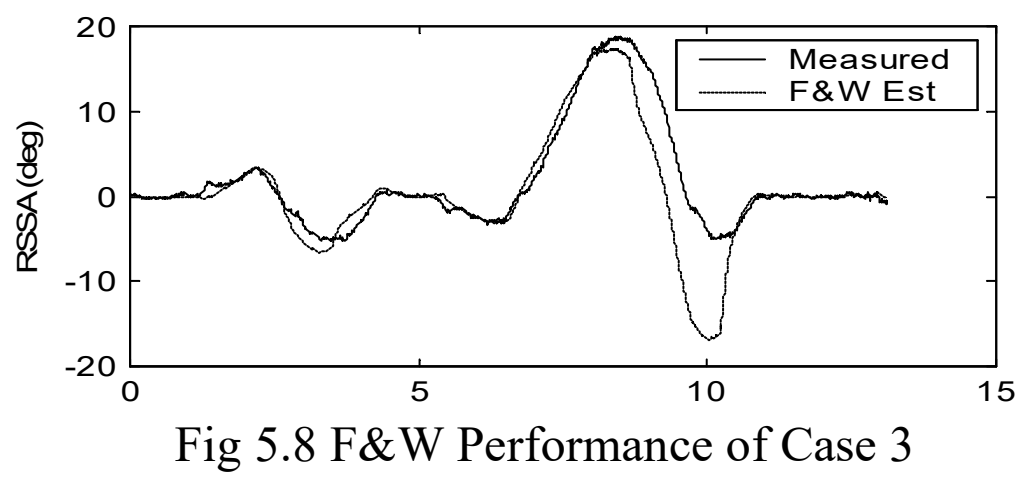

Fig 5.8 F&W Performance of Case 3

While the low-mu maneuvers inducing slow and sustained side slip angle or continuously sinusoidal side slip angle may be the most challenging ones for lateral velocity estimation, it is critical to have an observer that does not sacrifice its robustness during hi-mu maneuvers involving sensor noises, especially the ones induced by road disturbances.

Fig. 5.9 shows a banked slalom maneuver on a hi-mu asphalt surface. Fig. 5.10 demonstrates the robustness of SMO estimates under road bank disturbances while Fig. 5.11 reveals the weakness of F&W estimates under the same conditions. Note that the same compensation of lateral acceleration bias (shown in Fig. 5.9) has been performed for both methodologies.

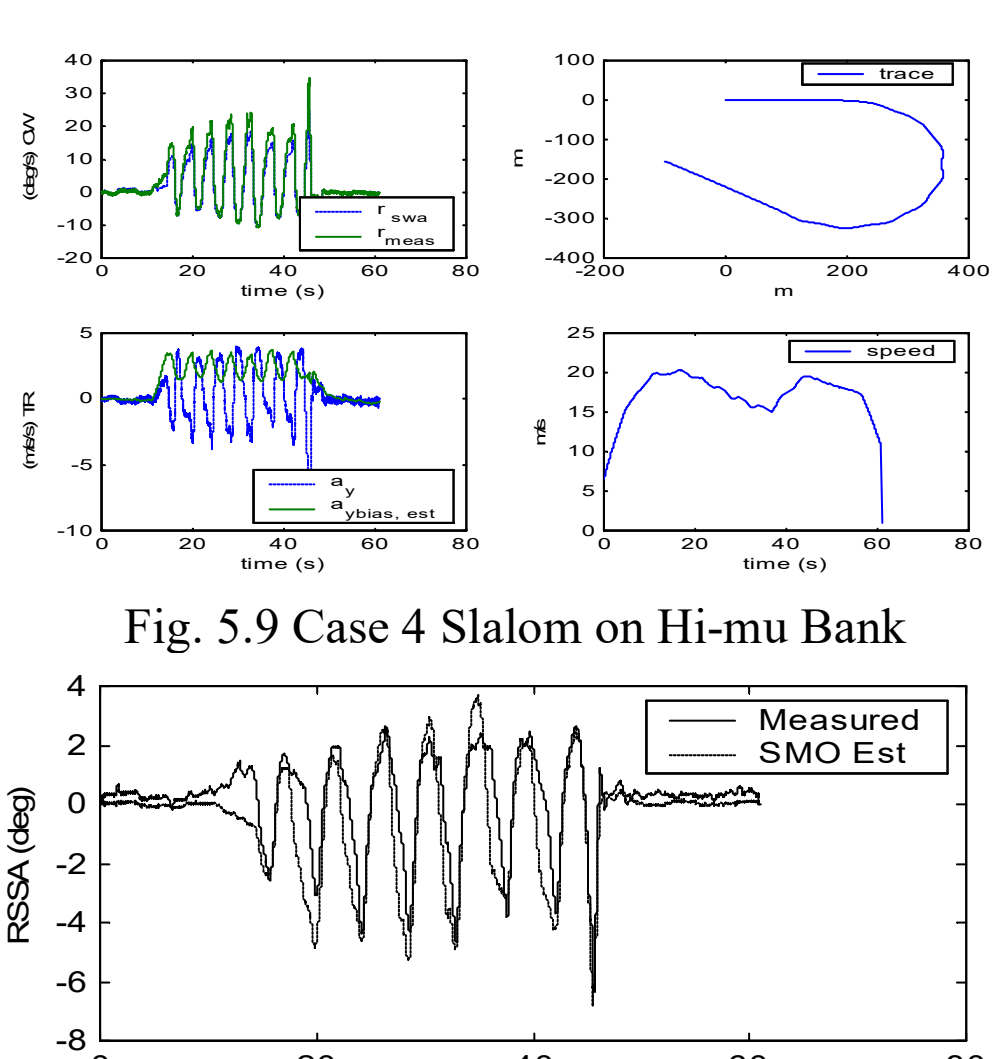

Fig. 5.9 Case 4 Slalom on Hi-mu Bank

Fig. 5.10 SMO Performance of Case 4

Fig. 5.11 F&W Performance of Case 4

## 6. CONCLUSIONS

In this paper, an adaptive sliding mode observer for lateral velocity estimation is developed. The effect of the design parameter on the base algorithm is studied through simulations. And the robust performance under a wide range of realistic maneuvers/environments is demonstrated through the post processing of experimentally collected data.

Most importantly, the mathematical stability of the devised observer is shown analytically and the observer performance is demonstrated during the more challenging maneuver (in terms of lateral velocity estimation) of a slow J-turn on low mu.

## ACKNOWLEDGEMENT

The author thank Dr. Davor Hrovat and Dr. Jing Sun of Ford Research Laboratory for their technical guidance and suggestions on this study.